\documentclass[twocolumn,showpacs,showkeys,preprintnumbers,amsmath,amssymb]{revtex4}
\usepackage{amsfonts}
\usepackage{amssymb}
\usepackage{color}
\usepackage{txfonts}
\usepackage{graphicx}
\usepackage{dcolumn}
\usepackage{bm}
\usepackage{cases}
\usepackage{amssymb}
\usepackage{txfonts}
\usepackage{graphicx}
\usepackage{dcolumn}
\usepackage{bm}
\usepackage{amssymb}
\usepackage{amsmath}
\usepackage{epsfig}
\usepackage{multirow}
\usepackage{threeparttable}
\begin{document}

\preprint{}
\title{Improvement on Corrosion Resistance and Biocompability of ZK60 Magnesium Alloy by Carboxyl Ion Implantation}
\author{Xian Wei}
\author{Su-Jie Ma}
\author{Pin-Duo Liu}
\author{Zhi-Cheng Li}
\author{Xu-Biao Peng}
\author{Rong-Ping Deng}
\author{Qing Zhao}
\email[The corresponding author Email: ]{qzhaoyuping@bit.edu.cn}
\affiliation{Beijing institute of technology}
\begin{abstract}
Magnesium alloys have been considered to be potential biocompatible metallic materials.
Further improvement on the anti-corrosion is expected to make this type of materials more suitable for biomedical applications in the fields of orthopedics, cardiovascular surgery and others.
In this paper, we introduce a method of carboxyl ion (COOH$^{+}$) implantation to reduce the degradation of ZK60 Mg alloy and improve its functionality in physiological environment.
X-ray photoelectron spectroscopy (XPS) and atomic force microscopy (AFM) experiments show the formation of a smooth layer containing carbaxylic group, carbonate, metal oxides and hydroxides on the ion implanted alloy surface.
Corrosion experiments and \emph{in vitro} cytotoxicity tests demonstrate that the ion implantation treatment can both reduce the corrosion rate and improve the biocompatibility of the alloy.
The promising results indicate that organic functional group ion implantation may be a practical method of improving the biological and corrosion properties of magnesium alloys.

\end{abstract}


\maketitle


\section{Introduction}
Because of their mechanical and biocompatible properties, magnesium alloys are promising materials for applications in medical fields \cite{catt2017poly, kirkland2012assessing}.
Mg alloys can provide effectively mechanical support for the injured tissue during healing period and can be completely biodegraded in the human body after being cured without a removal surgery \cite{jin2017corrosion, witte2008degradable}.
Magnesium alloys also present high strength to weight ratio and excellent mechanical properties.
The densities of magnesium alloys range from 1.74 to 2.0 g/cm$^{3}$ and the elastic moduli are from 41 to 45 GPa which are closer to the bones than that of the Ti-based alloys (4.4-4.5 g/cm$^{3}$ and 110-117 GPa) and stainless steels (7.9-8.1 g/cm$^{3}$ and 189-205 GPa) \cite{staiger2006magnesium}.
In addition, the released Mg ions are indispensable in many biological processes and have positive effect on the osteoblast adhesion and proliferation \cite{hickey2015adding, witte2005vivo, elin2010assessment}.
Despite possessing these promising characteristics, rapid and uncontrollable corrosion rate has restricted their applications in biomedical field.
The fast corrosion rate may lead to the failure of mechanical support, excessive corrosion products, and formation of local alkaline environment.
In order to improve the anticorrosion of Mg alloys, various methods have been proposed, including alloying, coating, and surface treatment such as anodization, physical/chemical vapor deposition.
However, these methods have drawbacks.
For example, the alloying method may improve the anticorrosion property by compromising the mechanical property and the biocompatibility, while the coating method may have inadequate bonding between the coating layer to the alloy \cite{bagherifard2018effects}.

The method of ion implantation is a surface treatment by introducing foreign ions to form a stable functional layer on the material surface in order to reduce the corrosion rate and improve the biocompatibility of magnesium alloys.
The energetic ions directly implanted into the ally surface do not change the size and shape of the material objects.
Besides, the implantation process is clean and environmentally friendly.
A number of literatures have been reported the positive effects of ion implantation on Mg alloys.
Tao et al. \cite{tao2014nanomechanical} grafted Gd into ZK60 Mg alloy by ion implantation and demonstrated that the corrosion resistance, surface hardness and modulus of the Gd implanted magnesium alloy are enhanced.
Liu et al. \cite{liu2017improved} revealed that the cytocompatibility of MC3T3-E1 cells is improved and the corrosion behavior is comparable after Zn ion implantation and deposition of the Mg-1Ca alloy.
Jamesh et al. \cite{jamesh2014effects} introduced Zr, O, and Zr $\&$ O ions into the ZK60 Mg alloy surface by plasma immersion ion implantation.
Their work showed the enhanced corrosion resistance of the treated alloy.
However, excessive amounts of metal ions in human body could cause toxic effects.
Li et al. had reported that cerium (Ce), praseodymium (Pr) and yttrium (Y) can induce rigorous hepatotoxicity when exceeding the allowable dosage \cite{li2013novel}.
Thus, the elements with no or low toxicity and larger tolerance dosage should be taken into consideration for ion implantation.


Organic functional carboxylic ions (COOH$^{+}$) are less toxic as compared to metallic ions and have presented excellent biosafety and biocompatibility.
As reported in the previous studies \cite{li2002cell, zhang2013influence}, COOH$^{+}$ implantation leads to much better cell attachment and proliferation on the surface of polypropylene and multiwalled carbon nanotubes.
Carboxylic ion implantation also improves the catalytic activity of indium tin oxide electrode which has many applications in new biosensors and biofuel cell \cite{li2010cooh+, li2009direct, gao2006determination}.
However, to the best of our knowledge, there is little report about the corrosion resistance and the biocompatibility of COOH$^{+}$-implanted Mg alloys.
In this paper, we implant the carboxylic ion into the ZK60 alloy surface with doses of $1\times10^{16}$ and $5\times10^{16}$ ions/cm$^{2}$ at the energy of 200 KeV.
We identify the surface chemical structure of treated samples by means of X-ray photoelectron spectroscopy (XPS) and atomic force microscopy (AFM).
The mechanical performance is measured by Nano indenter.
The corrosion property is tested by electrochemical and immersion assays.
And the biocompatibility is investigated by cell viability assay \emph{in vitro}.


\section{Experimental details}
\subsection{Ion implantation and surface characterization}
The ZK60 Mg alloy with the chemical contents of 5.5wt.\% Zn and 0.6wt.\% Zr was cut into $10mm\times10mm\times2mm$ chips.
Prior to ion implantation, samples were ground by SiC sandpapers up to 3000 grit and then rinsed with consecutive solutions of acetone, absolute ethanol, distilled water in ultrasonic cleaner, respectively.
The samples were finally dried in air.
Carboxy ion (COOH$^{+}$) implantation was carried out with the doses of $1\times10^{16}$ and $5\times10^{16}$ ions/cm$^{2}$ using an ion implanter that was equipped with formic acid feeding for ion source at an extraction voltage of 200 KV and a vacuum of $10^{-3}$ Pa.
COOH$^{+}$ ions were identified by mass spectrometry. The ions were implanted into the ZK60 sample with the ion current density less than 0.01 A/cm$^{2}$.

The X-ray photoelectron spectroscopy (XPS, PHI Quantera II, Ulvac-Phi Inc.) was used to determine the elemental depth profile and the surface chemical compositions of the treated samples.
The sputtering rate was approximately 25.8 nm/min based on the SiO$_{2}$ reference.
The morphology and roughness of the untreated and the implanted samples are characterized by Atomic force microscopy (AFM, MFP-3D-SA, Asylum Research, USA).
Additionally, Nano Indenter (XP, MTS Systems Corporation, USA) was used to study the change of surface mechanical properties of the samples after ion implantation.


\subsection{Electrochemical tests}
The electrochemical workstation was utilized to determine the electrochemical corrosion properties.
The tests were conducted at the temperature of 37$^{\circ}$C using a classical three electrode system with a platinum electrode as the counter electrode, a saturated calomel electrode (SCE) as the reference electrode, and the tested sample as the working electrode.
The sample with the exposed area of 1 cm$^{2}$ was soaked in Hank$^{\prime}$s solution.
The constituents of Hank$^{\prime}$s solution were NaCl 8.00 g/L, KCl 0.40 g/L, CaCl$_{2}$ 0.14 g/L, MgCl$_{2}$$\cdot$6H$_{2}$O 0.10 g/L, MgSO$_{4}$$\cdot$7H$_{2}$O 0.10 g/L, KH$_{2}$PO$_{4}$ 0.06 g/L, Na$_{2}$HPO$_{4}$$\cdot$2H$_{2}$O 0.06 g/L, NaHCO$_{3}$ 0.35 g/L, and glucose 1.00 g/L.
The electrochemical impedance spectroscopy (EIS) was performed with scanning frequency ranged from 100 kHz to 100 mHz at 5 mV amplitude perturbation.
And the potentiodynamic polarization tests were conducted at a scanning rate of 1 mV/s.
The corrosion current density ($I_{corr}$) were calculated by analyzing the Tafel extrapolation to the cathodic and anodic part of the polarization curves.

\subsection{Immersion tests}
The immersion tests were conducted in Hank$^{\prime}$s solution with volume to area ratio of 20 mL/cm$^{2}$ at $37^{\circ}$C depending on ASTM-G31-72 \cite{internasional2004astm}.
The weight loss, pH value, and ion concentration were recorded at each time node.
After immersion for 1, 3 and 7 days, the corroded surface was cleaned with chromic acid, gently rinsed by distilled water, and then dried in open air.
The releases of magnesium, calcium, and phosphate species were evaluated using inductively coupled plasma atomic emission spectrometry (ICP-OES, 725-ES, Agilent, USA).
The surface morphology of each corroded sample was analyzed by scanning electron microscopy (SEM, S-4800, Hitachi, Japan).


\begin{figure}[!htp]
\includegraphics[width=7.0cm]{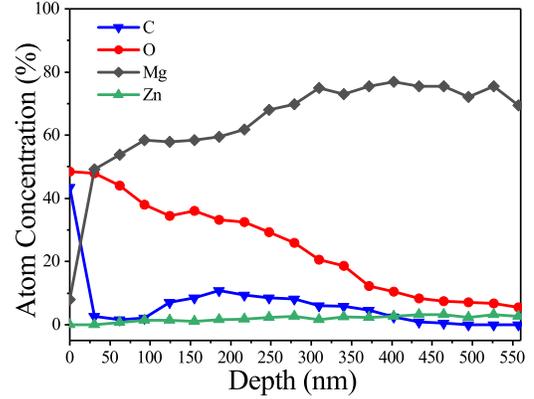}
\caption{\label{Fig1} The elemental depth profiles of the treated sample with dose of $5\times10^{16}$ COOH$^{+}$/cm$^{2}$ by XPS.}
\end{figure}

\begin{figure*}[!htp]
\includegraphics[width=13.0cm]{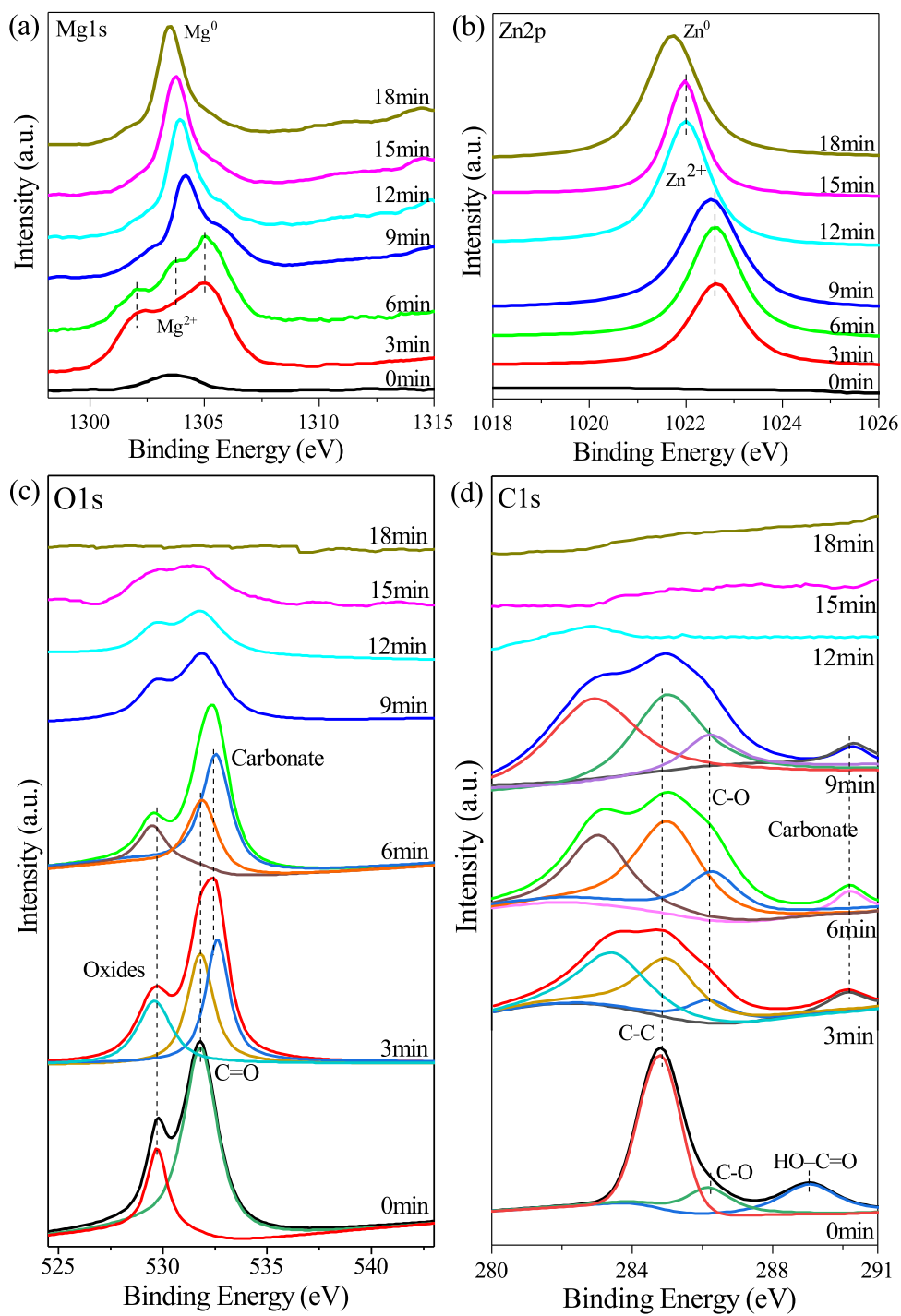}
\caption{\label{Fig2} High-resolution XPS (a) Mg 1s, (b) Zn 2p, (c) O 1s, and (d) C 1s spectra of the treated sample with dose of $5\times10^{16}$ COOH$^{+}$/cm$^{2}$ at different sputtering time.}
\end{figure*}


\subsection{\emph{In vitro} cytotoxicity tests}
The osteoblast MC3T3-E1 mice cells were acquired from American Type Culture Collection (ATCC) and used to test the negative effect on cell viability with an indirect method.
The cells were cultured in Dulbecco's Modified Eagle's Medium (DMEM, Gibco) along with 10\% fetal bovine serum (FBS) at 37$^{\circ}$C in a humidified atmosphere of 5\% CO$_{2}$.
Samples were sterilized under ultraviolet radiation for 2 hrs prior to the test.
The extracts were prepared by culturing the specimens in the serum free cell culture medium with the extraction ratio of 1.25 cm$^{2}$/ml in the incubator with 5\% CO$_{2}$ at 37$^{\circ}$C for 72 hrs.
The supernatant fluid was withdrawn, centrifuged, and then refrigerated at 4 $^{\circ}C$ for the test.
MC3T3-E1 cells were seeded in 96-well culture
plate with a density of $1 \times 10^{3}$ cells per well.
After the incubation for 24 hrs to allow attachment, the medium was replaced by the extracts from the untreated and the treated ZK60 Mg alloys adding with 10\% FBS.
The control group was normal DMEM medium with 10\% FBS.
After incubation for 1 and 3 days, each well was added by 10 $\mu$L MTT, and then cultured for another 4 hrs.
Then 150 $\mu$L dimethyl sulfoxide was added in each well to dissolve the formed formazan crystals.
The absorbance peak at 490 nm was observed by the microplate reader (Cytation3, Bio-Teh, USA).
In the end, the cell viability was computed as follows:
\begin {equation}
Viability = \frac{OD_{sample}}{OD_{control}}\times 100 \%
\end {equation}

\subsection{Statistical analysiy}
The data in this work expressed as mean $\pm$ standard deviations were analyzed by one-way analysis of variance (ANOVA).
Statistical analyses were operated using IBM SPSS Statistics 20.0 for Windows Software (IBM Inc., Armonk, USA) and $P<0.05$ reflected statistical significant.

\begin{figure*}[!htbp]
\centering
\includegraphics[width=17.1cm]{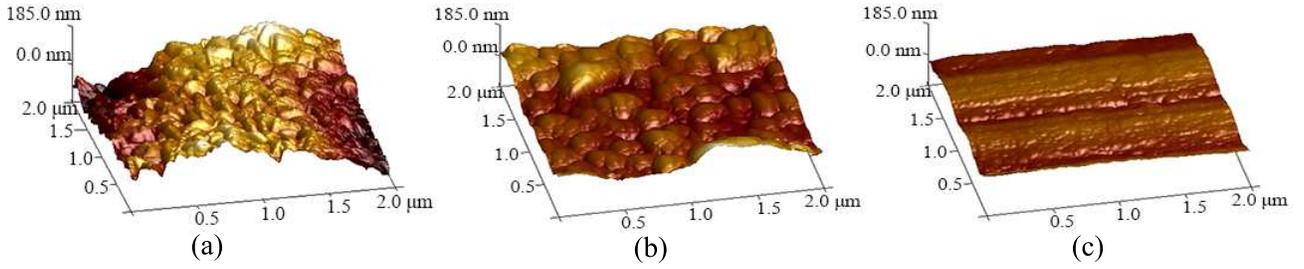}
\caption{\label{Fig3}  Surface morphology of the samples observed by AFM: (a) untreated sample, (b) implanted sample with dose of $1\times10^{16}$ ions/cm$^{2}$, and (c) implanted sample with dose of $5\times10^{16}$ ions/cm$^{2}$.}
\end{figure*}

\begin{figure*}[!htbp]
\includegraphics[width=14.0cm]{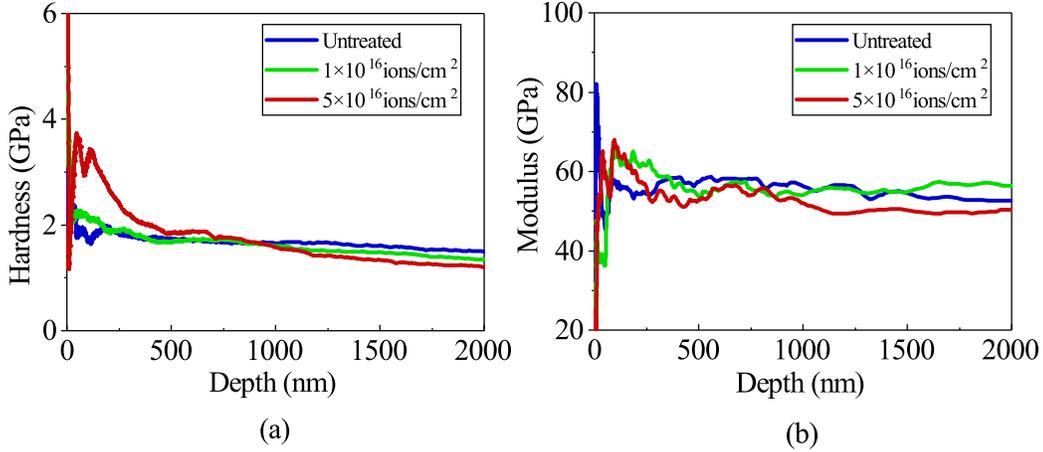}
\caption{\label{Fig10} The hardness (a) and elastic modulus values (b) acquired from the untreated and the treated samples.}
\end{figure*}

\section{Results}
\subsection{Surface characterization}
Fig. \ref{Fig1} shows the XPS depth profile of the treated sample which is implanted by COOH$^{+}$ ions with the dose of $5\times10^{16}$ ions/cm$^{2}$.
The magnesium concentration increases gradually with the sputtering depth while the reverse trend is found for oxygen.
The concentration of carbon is about 43.5\% when the sputtering begins, then drops sharply to 3\% at a depth of 25 nm.
This change may be due to the air-exposed surface of the sample.
Subsequently, the concentration of carbon increases to approximately 11\% with sputtering depth of nearly 185 nm.
Both oxygen and carbon concentrations decrease to low levels at the depth exceeding 465 nm, which indicates the formation of carboxylic-containing layer on the surface.
The high-resolution XPS spectra for Mg 1s, Zn 2p, O 1s, and C 1s regions of the treated sample with dose of $5\times10^{16}$ COOH$^{+}$/cm$^{2}$ are depicted in Fig. \ref{Fig2}.
All the peak positions are calibrated with the C1s adventitious peak (binding energy of 284.8 eV) \cite{ur2017effects}.
The Mg 1s spectrum is deconvoluted into three peaks at the beginning of sputtering, which signifies the formation of oxide, hydroxide, and carbonate in the top surface layer.
As the sputtering time continues, those three peaks gradually convert to one peak corresponding to MgO.
Finally the Mg 1s peak shifts to low binding energy, which indicates that Mg compounds turn steadily to the metallic state (Mg$^{0}$).
As shown in Fig. \ref{Fig2}(b), peak positions at 1022.0 eV and 1022.62 eV are assigned to the oxide and hydroxide of Zn respectively \cite{ballerini2007acid}.
The binding energy of Zn 2p peak exhibits a decreasing trend which indicates that Zn changes from its oxidized/hydroxy state to the metallic state.
The C 1s spectrum is deconvoluted into three peaks at the beginning of the sputtering time, including the adventitious carbon (C-C, 284.8 eV), the hydroxyl carbon (C-O, 286.17 eV), and the carboxylate carbon (HO-C=O, 289.0 eV), which indicates that the successful grafting of carboxylic functional group on the surface of ZK60 magnesium alloys.
Subsequently, the intensity of C changes significantly with increasing in depth.
The peaks near 290.0 eV is associated with the formation of carbonates, which is consistent with the Mg peak \cite{ni2008differentiating, christie1981xps}.
For the oxygen states, the O 1s spectrum is mainly separated into three peaks at the beginning of sputtering, which correspond to oxide, carbonate, and carbonyl groups, respectively.
The presence of carbon and oxygen confirms that the carboxylic functional group is successfully grafted on the alloy surface to form various compounds, such as carbonates, metal oxides and hydroxides.
The formation of the functional layer on top of the Mg substrate can provide to protect the alloy from corrosion by the external aggressive solution.

Fig. \ref{Fig3} shows the surface 3D topography of the untreated and the treated samples observed with atomic force microscopy (AFM).
The surface morphology of the untreated sample is quite rough with obvious convex.
The compact, uniform and relatively smooth surfaces form after the implantation of COOH$^{+}$ ions on the untreated sample.
The AFM images are taken with the area of $2 \times 2 \mu m$. The corresponding root-mean-square (RMS) roughness values are calculated.
The RMS value of the untreated sample is 52.8 nm, whereas RMS values are significantly lower for the treated samples (41.7 nm for the treated sample with dose of $1\times10^{16}$ ions/cm$^{2}$, 18.9 nm with dose of $5\times10^{16}$ ions/cm$^{2}$).
As the implantation dose increased from $1\times10^{16}$ to $5\times10^{16}$ ions/cm$^{2}$, the surface tends to be much smoother which could be attributed to the surface restruction at the sputtering effect of ion beam bombardment and the energy impact during the implantation process \cite{xu2014eelectrochemical, jamesh2013effects, wan2007corrosion}.
Most previous studies have reported that evident smooth surface is observed after the ion implantation \cite{zhao2014enhanced, pandey2017surface, bolduc2001deep}.
Jamesh et al. demonstrated that the WE43 Mg alloy surface is smoother after Si ion implantation \cite{jamesh2013effects}.
Wan et al. also indicated that the surface of sample treated by oxygen ion implantation appears more homogeneous and smoother \cite{wan2007corrosion}.
The effect of carboxylic ion implantation on the surface morphology is obvious, which also has impact on the corrosion properties of the ZK60 Mg alloy.
Generally, the potential difference between the concave and convex is less for the smoother surface, which corresponds to more resistant to corrosion \cite{li2006influence, hong1997effect, sasaki1996generation}.
Therefore, the corrosion rate of ZK60 Mg alloy is expected to be reduced after COOH$^{+}$ ions implantation.


\subsection{Mechanical property}

The measurement results of the hardness and the elastic modulus for the untreated and the ion implanted samples are shown in Fig. \ref{Fig10}.
There are no significant changes in the hardness and the modulus from the outer layer to the inner layers on the untreated samples, while the surface up to about 465 nm on the ion implanted samples show higher values of the hardness and the elastic modulus.
The hardness and elastic modulus of all the treated samples are larger than those of the bare plate on the top layer and then tend to be stable in deeper indentation depth.
This suggests that ion implantation improves the mechanical performance of the ZK60 magnesium alloy \cite{poon2005carbon, zhao2012nano}.


\begin{figure}[!htbp]
\includegraphics[width=7.0cm]{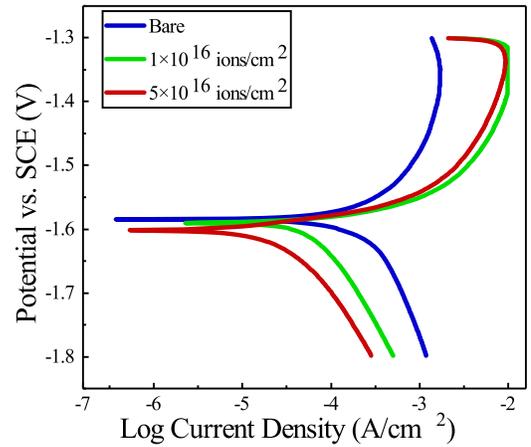}
\caption{\label{Fig6} The potentiodynamic polarization curves of the untreated and treated samples in Hank$\prime$s solution.}
\end{figure}

\begin{figure*}[!htbp]
\includegraphics[width=16.0cm]{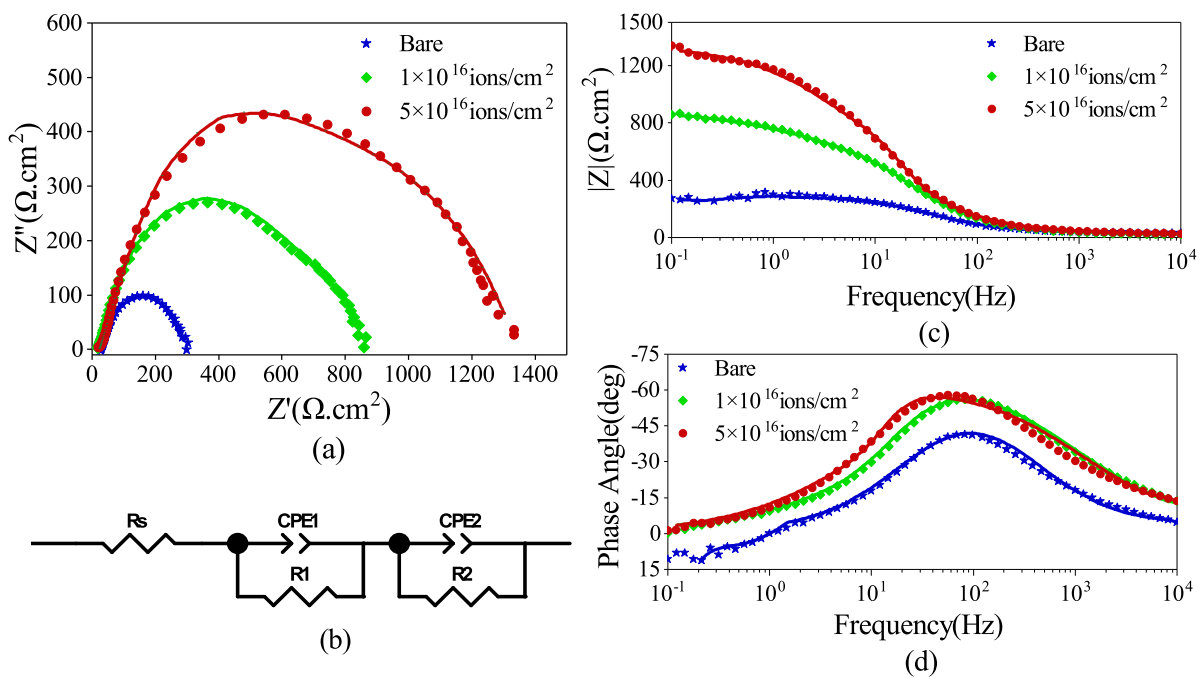}
\caption{\label{Fig5} EIS curve (scatter plot) and model fitting curve (solid line) of samples: (a) Nyquist plot, (b) Equivalent circuit of samples with and without implantation, (c) Bode plot of impedance versus frequency, and (d) Bode plot of phase angle versus frequency.}
\end{figure*}

\subsection{Corrosion performance}


Fig. \ref{Fig6} shows the potentiodynamic polarization measurement results for samples soaked in Hank$^{\prime}$s solution at $37^{\circ}$C.
The electrochemical parameters such as corrosion potential (E$_{corr}$) and corrosion current density (I$_{corr}$) are obtained by the Tafel extrapolation and listed in Table \ref{Table1}.
The I$_{corr}$ attributes lower values for the treated samples compared with the untreated one.
The corrosion current density of the implanted samples decreases with increasing the COOH$^{+}$ ion implantation dose.
The I$_{corr}$ of untreated sample displays nearly 9 times higher than the treated sample with $5\times10^{16}$ ions/cm$^{2}$, 6 times larger than the sample with $1\times10^{16}$ ions/cm$^{2}$.
Generally, more negative corrosion current density corresponds to lower corrosion rate \cite{xie2010control, chen2011deposition}.
The results of potentiodynamic polarization reveal that the samples have better anti-corrosion performance after COOH$^{+}$ ions implantation.
This can be attributed to the formation of functional layer containing carboxylic group, carbonates, metal oxides and hydroxides, which provides a barrier to reduce the corrosion of ZK60 magnesium alloy.

\begin{table}
\caption{\label{Table1} Corrosion parameters calculated from the potentiodynamic polarization curves.}
\setlength{\tabcolsep}{1 mm}{
\begin{tabular}{|c|c|c|} 
\hline
Doses(ions/cm$^{2}$) & E$_{corr}$(V) & i$_{corr}$(A/cm$^{2}$)\\
\hline
 0 & -1.585 & $2.9194\times10^{-4}$ \\
 $1\times10^{16}$ & -1.59 & $4.9799\times10^{-5}$ \\
 $5\times10^{16}$ & -1.601 & $3.3245\times10^{-5}$\\
 \hline

 \end{tabular}}
\end{table}

\begin{figure*}[!htbp]
\includegraphics[width=14.0cm]{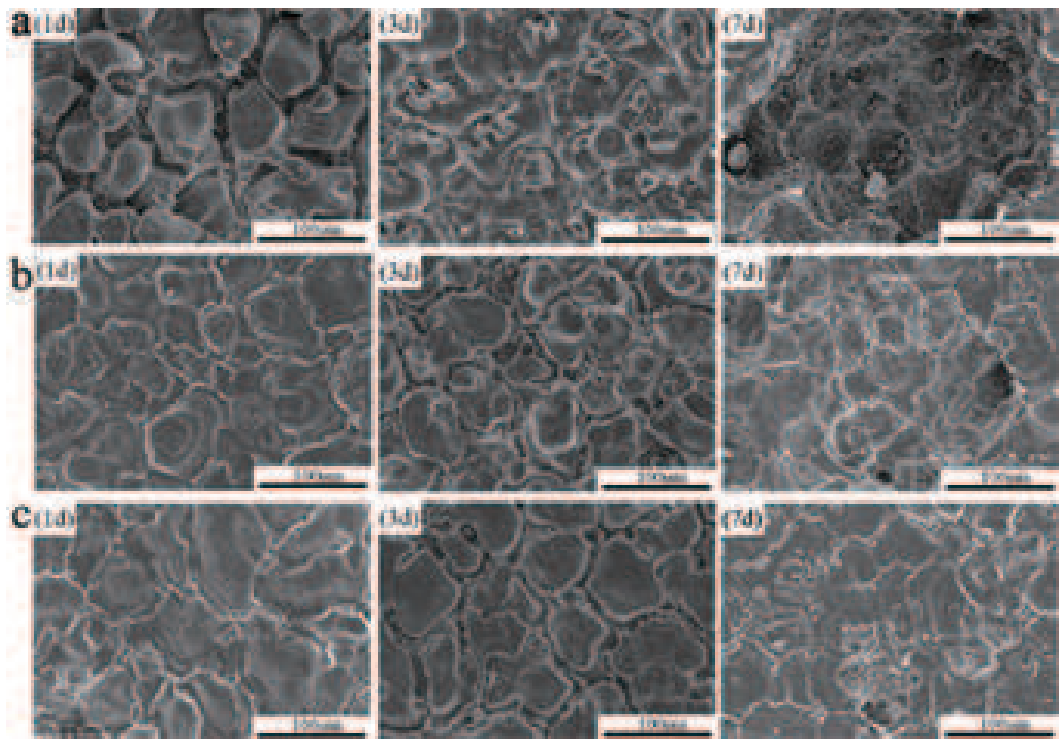}
\caption{\label{Fig7} Surface morphology of samples during different immersion time. Group a the untreated sample, group b the treated sample with dose of $1\times10^{16}$ ions/cm$^{2}$, and group c the treated sample with dose of $5\times10^{16}$ ions/cm$^{2}$.}
\end{figure*}

\begin{figure*}[!htbp]
\includegraphics[width=14.0cm]{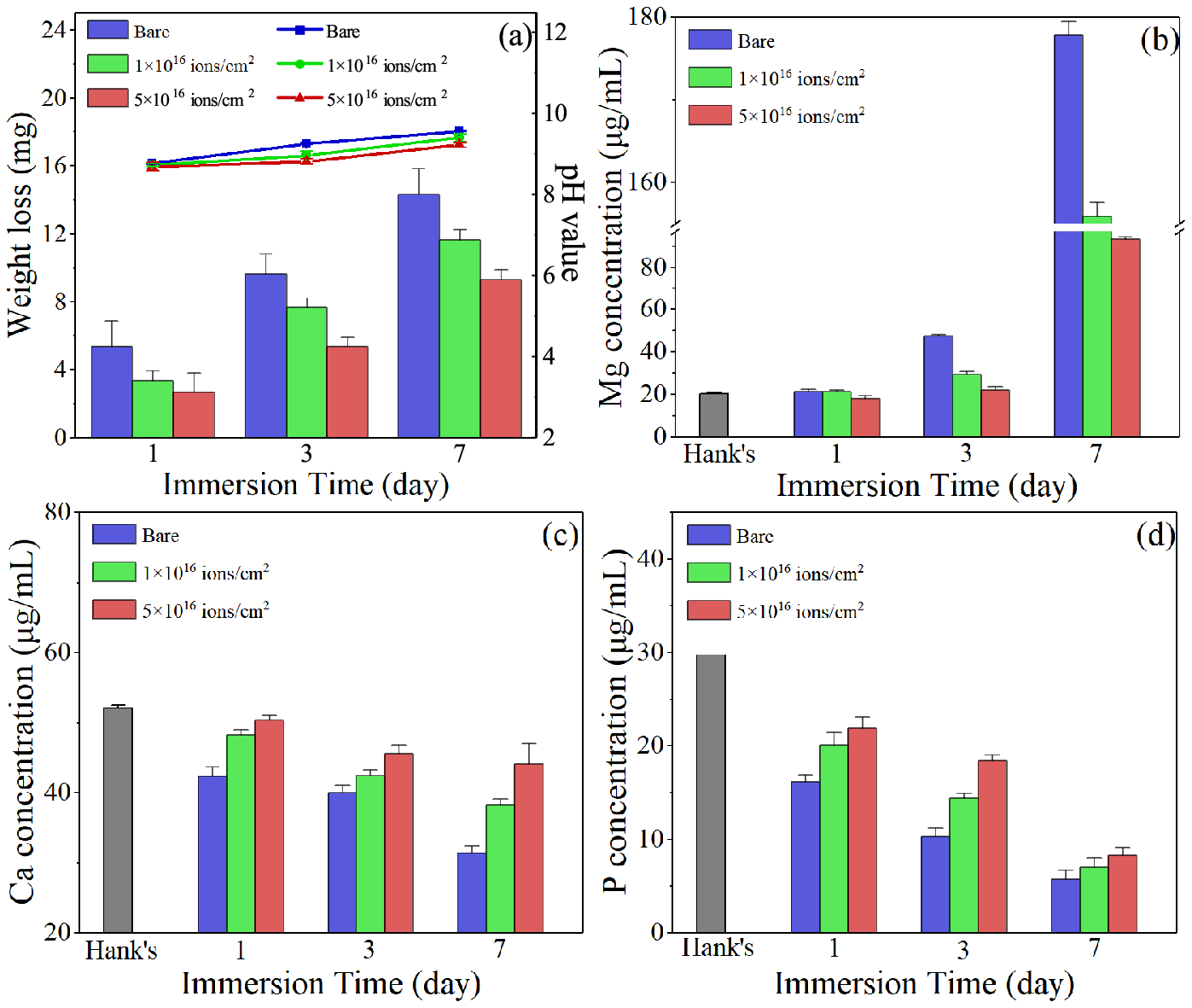}
\caption{\label{Fig8} (a) The pH values extracted from the Hank's solution and the weight loss after immersion for 1, 3, 7 days. And the concentrations of (b) Mg, (c) Ca, and (d) P after immersion for 1, 3, 7 days.}
\end{figure*}

The electrochemical impedance spectroscopy (EIS) is also employed to analyze the corrosion property of the electrode system.
Fig. \ref{Fig5} shows the corresponding Nyquist plot and Bode plots, including impedance versus frequency and phase angle versus frequency.
As shown in Fig. \ref{Fig5}(a), two capacitive loops are observed at high and medium frequencies.
They are distinguishable for the treated sample, while almost overlapped for the untreated sample.
The diameter of the capacitive semicircle is significantly larger with the dose of $5\times10^{16}$ COOH$^{+}$/cm$^{2}$ than with the dose of $1\times10^{16}$ COOH$^{+}$/cm$^{2}$.
For the untreated sample, the diameter becomes very small.
As reported in literatures \cite{singh2015comparative, ruhi2009corrosion}, capacitive loop at high frequency region is related to the property of the coating and the charge transfer process, and the capacitive loop at low frequency region is associated with the mass transfer process.
The formations of carboxylic group, carbonates, metal oxides and hydroxides on the surface of the treated sample act as coating to protect the ZK60 alloy from corrosion.
Therefore, the anti-corrosion performance is enhanced for the COOH$^{+}$ ion implanted ZK60 Mg alloy.


The Bode impedance plots of the untreated and the treated samples are shown in Fig. \ref{Fig5} (c).
The impedances of the treated samples are all higher than the untreated sample at the lower frequency regions, whereas they are almost indistinguishable in higher frequency regions.
At the frequency of 100 mHZ, measured impedance value of the treated sample with dose of $5\times10^{16}$ ions/cm$^{2}$
is the impedance five times higher than that of the untreated sample, and three times higher impedance for the treated sample with dose of $1\times10^{16}$ ions/cm$^{2}$.
A survey of the literatures \cite{chen2007long, xu2014eelectrochemical, amirudin1995application} shows that the higher impedance at low frequency (100 mHz) reflects slower charge transfer process.
Fig \ref{Fig5} (d) shows the Bode phase angle evolution process of the untreated and treated sample.
The maximum phase angles of the samples are -41.5 degrees (bare sample), -56.2 degrees (treated sample with dose of $1\times10^{16}$ ions/cm$^{2}$), and -57.9 degrees (treated sample with dose of $5\times10^{16}$ ions/cm$^{2}$), respectively.
More capacitive behavior is demonstrated by larger phase angles of the treated samples, which suggests the formation of a more stable and dense layer to retard the electrolyte penetration into the substrate \cite{murray1997electrochemical, shi2005performance, park2002anticorrosive, wang2014effect, de2006corrosion}.
Hence, the Bode diagram confirms that corrosion resistance of the alloy is improved by the carboxyl ion implantation.


The electrical equivalent circuit is employed to fit the EIS spectra (Fig. \ref{Fig5} (b)).
The fitted results are plotted as solid lines throughout the Nyquist and Bode plots as shown in Fig. \ref{Fig5} (a, c, d).
Their corresponding EIS parameters derived from curve fitting are listed in Table \ref{Table2}.
R$_{s}$ is the solution resistance between the working electrode and reference electrode.
CPE$_{1}$ represents the constant phase element of the outer porous layer, and R$_{p}$ is the corresponding resistance.
CPE$_{2}$ stands for the constant phase element in the inner layer and in paralleled with the resistance R$_{2}$.
From Table \ref{Table2}, R$_{s}$ maintains relatively unchanged for different compositions of the samples in the same solution.
As reported that the CPE value is related to the ability of the electrolytes to permeate into the porous surface layer \cite{singh2009influence, bellucci1993water}.
A decrease of CPE suggests that it is difficult to pass through the porous/defective surface layer for electrolytes.
The values of R$_{1}$ and R$_{2}$ increase with the decreasing of CPE$_{1}$ and CPE$_{2}$ for the treated samples compared to the untreated sample, which indicates the lower penetration rate of the Hank$^{\prime}$s solution into the treated sample surface.
Moreover, R$_{1}$ has a significantly lower value than R$_{2}$ for the untreated sample.
It implies the formation of the thin porous outer layer, which can not suppress the transfer of the corrosive chloride ions and other chemical compounds.
So the magnesium substrate of the untreated sample is easily attacked by the abundant chloride ions.
On the other hand, the significantly higher values of $R_{1}$ and $R_{2}$ of the treated samples are due to the formation of carboxylic group, carbonates, metal oxides and hydroxides on the surface layer after carboxyl ion implantation.
Thus the formed compact and protective structure on the surface of the treated sample possess a barrier to protect the ZK60 Mg alloy from corrosion.
The above EIS results suggest that the treated samples effectively improve the corrosion resistance and are well consistent with the polarization curve.

\begin{table}
\caption{\label{Table2} Electrochemical parameters derived from the equivalent circuit models.}
\setlength{\tabcolsep}{0.3 mm}{
\begin{tabular}{|c|c|c|c|} 
\hline
 Parameters & Bare & $1\times10^{16}$ &$5\times10^{16}$ \\
 \hline
 $R_{s}$ ($\Omega cm^{2}$) & 20 & 19.363 & 19.12 \\
 $R_{1} (\Omega cm^{2}$) & 15.849 & 360.8 & 338.9 \\
 $R_{2} (\Omega cm^{2})$ & 253.47 & 496.68 & 950.49 \\
 $CPE_{1}-T$ ($\Omega ^{-2}cm^{2}S^{-n}$) & $5\times10^{-4}$ & $4.8789\times10^{-4}$ & $4.3942\times10^{-4}$ \\
 $CPE_{1}-P$ & 0.363 & 0.63736 & 0.8271 \\
 $CPE_{2}-T$ ($\Omega^{-2}cm^{-2}S^{-n}$) & $5.1949\times10^{-5}$ & $3.1653\times10^{-5}$ & $3.0092\times10^{-5}$ \\
 $CPE_{2}-P$ & 0.85261 & 0.88782 & 0.84661 \\
  \hline

 \end{tabular}}
\end{table}

To further evaluate the corrosion behavior, the corrosion morphologies of the untreated and treated samples are analyzed with SEM, as shown in Fig. \ref{Fig7}.
The samples were soaked in the Hank$^{\prime}$s solution at 37$^{\circ}$C for 1, 3, and 7 days, respectively.
Plenty of cracks are viewed on the surface for the untreated sample, and the crack spacing propagates obviously with the immersion time, as shown in Group a in Fig. \ref{Fig7}.
Severe large corrosion pits emerge on the untreated sample surface after immersion for 7 days, which specifies its highest corrosion rate.
On the other hand, for the treated samples with dose of $5\times10^{16}$ ions/cm$^{2}$, except some eroding cracks and pits, their surfaces haven't undergone visible damage.
The treated samples display smaller crack spacing compared with untreated sample after immersion for 7 days.
Thus, the corrosion of the ZK60 Mg alloy is effectively suppressed by the COOH$^{+}$ implantation on the surface.

Additionally, Fig. \ref{Fig8} shows the weight loss, the pH value measured from the Hank$'$s solution, and the released ion concentrations of Mg, Ca, P of the untreated and treated samples.
As shown in Fig. \ref{Fig8} (a), the tendency of the continuous increase of the weight loss of the samples is observed with longer immersion time.
The untreated sample exhibits larger weight loss compared with the treated sample at each time node.
For the treated samples, the weight loss decreases with increasing the implantation dose, indicating the superior corrosion resistance for the treated sample with higher dose.
Similar case is observed in the pH values of the untreated and treated samples.
The untreated sample has slightly higher pH values, which suggests a higher corrosion rate.
As shown in Fig. \ref{Fig8} (b-d), the release of Mg ion increases gradually over the immersion time.
The treated sample dissolves less Mg ion in the corrosive solution than the untreated sample.
The lowest dissolution of Mg is found for the sample implanted by the COOH$^{+}$ with dose of $5\times10^{16}$ ions/cm$^{2}$.
After immersion for 7 days, the concentrations of Ca and P in the solution extracted by the untreated sample are $31.33\pm1.04$ $\mu$g/mL and $5.73\pm0.95$ $\mu$g/mL, lower than the treated sample with dose of $5\times10^{16}$ ions/cm$^{2}$ (Ca: $44.10\pm2.92$ $\mu$g/mL and P: $0.27\pm0.83$ $\mu$g/mL).
The results are attributed to the more insoluble Ca and P corrosion products formed for the untreated sample in the corrosion process.
Therefore, the degradation rate of the treated sample is lower than the untreated sample.
These results are in consistent with the analyses of the surface morphology, and all the corrosion assays indicate that the corrosion behavior is effectively enhanced after the carboxyl ion implantation.

\begin{figure*}[!htbp]
\includegraphics[width=14.0cm]{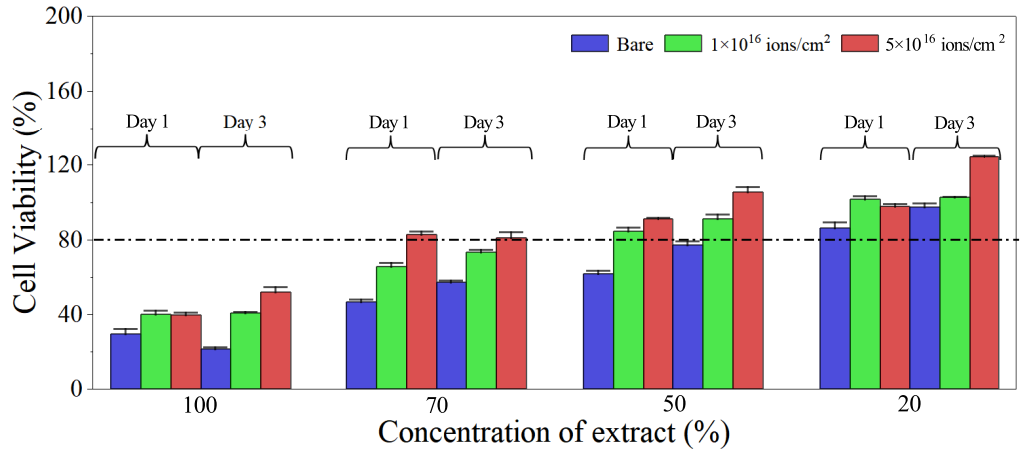}
\caption{\label{Fig9} \emph{In vitro} viability of MC3T3-E1 cells cultured in the different concentration extracts of bare and treated samples for 1 and 3 days.}
\end{figure*}

\subsection{\emph{In vitro} cytotoxicity studies}
The \emph{in vitro} cytotoxicity of the untreated and the treated ZK60 Mg alloys are evaluated by measuring the viability of MC3T3-E1 cells after culturing for 3 days in the extracts with different concentrations (20\%, 50\%, 70\%, and 100\%).
In the light of ISO 10993-5 \cite{ISO109, tong2018microstructure}, if the cell viability exceeds 80\%, the biological material exhibits none or slight cytotoxicity; if the cell viability ranges from 50\% to 79\%, it exhibits mild cytotoxicity; if the cell viability is between 30\% and 49\%, it exhibits moderate cytotoxicity; if the cell viability is below 30\%, it exhibits severe cytotoxicity.
As shown in Fig. \ref{Fig9}, both the untreated and the treated samples have relatively low cell viability in the extract with concentration of 100\% after being incubated for 3 days.
The cell viability of the untreated sample does not reach more than 80\% until the extract concentration reduces to 20\%, suggesting the severe toxicity effect.
However, the cell viability increases gradually as reducing the extract concentration.
The cell viability has grown to over 80\% for the implanted alloy with dose of $5\times10^{16}$ ions/cm$^{2}$ when the extract concentration is 70\%, showing a good cytocompatibility.
All the treated samples with different implantation doses achieve high cell viability above 80\% when the extract concentration is less than 50\%.
Additionally, the cell viability of the treated sample with dose of $5\times10^{16}$ ions/cm$^{2}$ is visibly larger than that of the dose of $1\times10^{16}$ ions/cm$^{2}$.
This is caused by that the treated sample with dose of $5\times10^{16}$ ions/cm$^{2}$ has lower degradation rate making for less ion dissolution in the extract solution.
Overall, the MC3T3-E1 cells grown in the extract of treated samples present significantly higher activity than that of the untreated sample for the whole extract concentration during the whole incubation period, indicating that the cytotoxic effect is restrained owing to the ion implantation into the alloy surface.




\section{Discussion}

\subsection{Corrosion behavior}

The magnesium corrosion mechanism has been reported in literatures \cite{song2003understanding, song1998influence, doepke2012corrosion}, the degradations of the Mg alloys involve the electrochemical reactions, which correspond to the metal dissolution at the anodic, and the oxygen reduction at the cathodic, as shown in the following chemical equations:

\begin{equation}
\begin{aligned}
Mg \rightarrow Mg^{2+}+2e^{-} \quad \text{(anodic reaction)}
\label{eq:11}
\end{aligned}
\end{equation}

\begin{equation}
\begin{aligned}
H_{2}O + 2e^{-} \rightarrow H_{2}+2OH^{-} \quad \text{(cathodic reaction)}
\label{eq:22}
\end{aligned}
\end{equation}

\begin{equation}
\begin{aligned}
Mg^{2+} + 2OH^{-} \rightarrow Mg(OH)_{2} \quad \text{(product formation)}
\label{eq:33}
\end{aligned}
\end{equation}

\begin{equation}
\begin{aligned}
Mg + 2H_{2}O \rightarrow Mg(OH)_{2} + H_{2} \quad  \text{(overall reaction)}
\label{eq:44}
\end{aligned}
\end{equation}

With respect to the untreated sample, magnesium dissolves rapidly in Hank$^{\prime}$s solution filled with plenty of chloride ions, which results in the increase of Mg$^{2+}$ concentration at anodic electrode.
Simultaneously, OH$^{-}$ ions are accumulated at cathodic electrode where the local high alkalinity is produced (Eq. (\ref{eq:22})).
The low soluble Mg hydroxide formed by the reaction of Mg$^{2+}$ and OH$^{-}$ (Eq. (\ref{eq:33})) precipitates on the alloy surface to serve as the initial barrier against the solution penetration.
However, this Mg(OH)$_{2}$ layer is too loose, porous, and unstable to protect the substrate when encountering with abundant chloride ions in the solution.
Then the non-uniform Mg(OH)$_{2}$ is converted into MgCl$_{2}$ further lessening the protection.
Thus, the Mg substrate underneath the porous corrosion product is readily corroded by the electrolyte.
The increase in corrosion current density and the decrease in EIS nyquist arc suggest that the porous surface layer is unable to stop the electrolyte permeation on the untreated sample.
As the corrosion progresses, the Mg phosphate and carbonate are formed by consuming the Ca and P ions in the corrosion solution.
The degradation of the ZK60 alloy leads to the increased amount of Mg ions and the decreased Ca and P ions in the solution.
The results of higher Mg concentration and lower Ca and P concentrations are shown in Fig. \ref{Fig8}, which indicates the impaired corrosion resistance of the untreated sample.

However, for the treated sample, the formation of the stable barrier layer containing carboxylic group, Mg carbonates, metal oxides and hydroxides effectively retards the electrolyte penetration into the alloy.
Both the higher R$_{1}$ and R$_{2}$ resistance values in the EIS analysis confirm the improved corrosion resistance of the treated sample.
Additionally, the R$_{2}$ is significantly higher than R$_{1}$, implying that the corrosion resistance is indeed dominated by the barrier layer formed by the carboxylic ion implantation.
It is noted that most of the surface areas are well protected by the barrier layer, while corrosion in some regions can be identified by examining the surface morphology (Fig. \ref{Fig7}), which can be explained as that the barrier layer is not homogeneously covers the whole alloy surface.

The concentration change by ICP-OES support the corrosion resistance.
The corrosion products, such as the Mg carbonate and phosphate, are are formed by consuming Ca and P ions.
Hence, the Mg concentration rises, whereas Ca and P concentrations decline, which verifies the formation of local alkaline environment.
In comparison with the bare sample, the treated sample exhibits lower Mg concentration and higher Ca and P concentrations, which reveals the reduced degradation rate.
Overall, the corrosion rate is significantly reduced after the grafting of carboxyl group on the surface of ZK60 Mg alloy.

\subsection{Cytotoxicity evaluation}
The mouse osteoblast MC3CT3 cells are used to evaluate the effect of the carboxyl ion implantation on cytotoxicity into the ZK60 Mg alloy.
The Mg alloy can gradually dissolves in the physiological environment after being implanted into the living body as the osteosynthesis material.
The degradation process leads to the increase in the pH value generating the local alkaline environment \cite{ohtsuka2012corrosion}.
The obvious cytotoxic effect is occurred when the amount of  the released ions from the implanted objects are beyond the maximum dosage tolerated by the tissues.
As reported in literature \cite{wong2013low}, the viability of MC3T3-E1 cells is significantly facilitated when the extract medium contains 50 ug/ml Mg ions, whereas the reduced viability is observed when the Mg ion concentration exceeds 200 ug/ml.
In this study, for 3 day culture period, the cell viability is smaller than 30\% for the untreated sample with extract at 100\%, encountering severe toxic effects as shown in Fig. \ref{Fig9}.
The cell viability keeps at a low level although there is an increase trend with diluted extract.
This implies that the high degradation rate of the untreated sample produces the high Mg concentration above the tolerance level, and in this alkaline environment, hydrogen gas accumulates surrounds the host tissue, leading to the necrosis and separation of tissues \cite{song2007control, staiger2006magnesium}.
Some studies have reported the similar results about the presence of cytotoxic effects of Mg-Zn alloys.
Jin et al. \cite{jin2017corrosion} demonstrated that the extremely low viability of the MC3T3-E1 pre-osteoblasts was found in 70\% extract of ZK60 Mg alloy.
Hong et al. \cite{hong2013vitro} also reported that the ZK40 Mg alloy possesses almost tiny cell viability with undiluted extract.
On the other hand, the treated samples have not induced obvious toxicity to MC3T3-E1 cells cultured with 70\% extract after the incubation for 3 days.
This demonstrates that the reduced corrosion rate leads
to the limited alkalization and hydrogen evolution to slow the release of ions from the Mg implants.
In addition, the carboxyl functional group facilitates the cell attachment and proliferation due to its less toxic than metal ions \cite{li2002cell, zhang2013influence}.
The carboxylic ion implantation which promotes the formation of  a relatively smooth and functional surface layer containing carboxylic group, carbonates, metal oxides and hydroxides on the substrate surface has shown obvious advantage in corrosion resistance, biocompatibility, and mechanical behavior.
All of these are vital factors for materials in clinical applications.


\section{Conclusions}
In this study, ZK60 magnesium alloy is implanted by carboxylic ions with the doses of $1\times10^{16}$ and $5\times10^{16}$ ions/cm$^{2}$ at energy of 200 KeV.
The performance of mechanical behavior, degradation rate and \emph{in vitro} cytotoxicity are systematically investigated.
The surface mechanical performance of surface modified alloy is improved after the ion implantation, which is confirmed by the Nano Indenter experiment.
The treated samples exhibit the reduced corrosion rate which is demonstrated by the experimental results in the potentiodynamic polarization, electrochemical impedance spectroscopy, and the immersion analyses.
The improvement can be attributed to the smoother surface layer containing various carbonates, metal oxides and hydroxides compound as a passive barrier layer to prevent the penetration of the corrosive solution.
The \emph{in vitro} cytotoxicity experiment using osteoblast MC3CT3 cell shows the higher viability of the treated samples with the promising biocompatibility.
Although this is only a preliminary study on the grafting of carboxyl onto magnesium alloys,
this study provides a perspective of employing the organic functional groups implantation techniques to improve the biodegradable magnesium alloys.

\section*{ ACKNOWLEDGMENTS }
This work is supported by National Science Foundation (NSF) of China with the Grant No.$11675014$. Additional support is provided by Ministry of Science and Technology of China $(2013YQ030595-3)$.

\newpage
\bibliographystyle{plain}

\bibliography{Reference}
\end{document}